\documentclass{elsarticle}

\usepackage{amssymb, amsmath}
\usepackage{graphics, array, url, hyperref, epsfig, amsmath,amssymb}
\usepackage{graphicx}
\usepackage{subfigure}
\usepackage{CJK}
\usepackage{latexsym,amsfonts}
\usepackage{tabularx}
\usepackage{amsmath}
\usepackage{epstopdf}
\usepackage{fancybox}
\usepackage{multirow}
\usepackage{array}
\usepackage[inline,draft,nomargin]{fixme}

\begin{document}

\begin{frontmatter}



\title{Blockchain based Privacy-Preserving Software Updates with Proof-of-Delivery for Internet of Things}

\author[snnu]{Yanqi Zhao}
\author[thirty]{Yiming Liu}
\author[snnu]{Yong Yu \corref{correspondingauthor1}}
\cortext[correspondingauthor1]{Corresponding author}
\ead{yuyong@snnu.edu.cn}
\author[uow]{Yannan Li}



\address[snnu]{School of Computer Science, Shaanxi Normal University, Xi'an, 710062, China.}

\address[thirty]{Science and Technology on Communication Security Laboratory, Chengdu 610041, China.}
\address[uow]{School of Computing and Information Technology, University of Wollongong, Wollongong, NSW 2522, Australia.}
\begin{abstract}
  A large number of IoT devices are connected via the Internet. However, most of these IoT devices are generally not perfect-by-design even have security weaknesses or vulnerabilities. Thus, it is essential to update these IoT devices securely, patching their vulnerabilities and protecting the safety of the involved users. Existing studies deliver secure and reliable updates based on blockchain network which serves as the transmission network. However, these approaches could compromise users privacy when updating the IoT devices.

 In this paper, we propose a new blockchain based privacy-preserving software updates protocol, which delivers secure and reliable updates with an incentive mechanism, as well protects the privacy of involved users. The vendor delivers the updates and it makes a commitment by using a smart contract to provide financial incentive to the transmission nodes who deliver the updates to the IoT devices. A transmission node gets financial incentive by providing a proof-of-delivery. The transmission node uses double authentication preventing signature (DAPS) to carry out the fair exchange to obtain the proof-of-delivery. Specifically, the transmission node exchanges an attribute-based signature from a IoT device by using DAPS. Then, it uses the attribute-based signature as a proof-of-delivery to receive financial incentives. Generally, the IoT device has to execute complex computation for an attribute-based signature (ABS). It is intolerable for resource limited devices. We propose a concrete outsourced attribute-based signature (OABS) scheme to resist the weakness. Then, we prove the security of the proposed OABS and the protocol as well. Finally, we implement smart contract in Solidity to demonstrate the validity of the proposed protocol.
\end{abstract}

\end{frontmatter}

\section{Introduction}\label{sec1}
According to Gartner Inc.\cite{gartner}, the IoT devices are deployed and connected on the Internet have more than 11 billion in 2018. 
  IoT and its applications have pervaded in our daily lives from smart home, smart city to smart everything.
   However, most of these IoT devices are generally not perfect-by-design with security weaknesses or vulnerabilities and are easy to be hacked under various cyber attacks. 
 In September 2018, ZeroDayLab \cite{zero} reports a high-severity vulnerability in the 4G-based wireless 4GEE Mini modem. The vulnerability could allow an attacker to run a malicious program on a targeted computer with the highest level of privileges in the system. Later, Mobile operator EE acknowledged the issue and rolled out a firmware patch to address the vulnerability. By using a previously disclosed vulnerability revealed in the CIA Vault 7 leaks, the hackers have compromised more than 210,000 routers from Latvian network hardware provider Mikrotik across the world, with the number still increasing \cite{mikrotik}, \cite{mikrotik1}. With the continues growth of IoT devices, it is essential to update these IoT devices securely, patching their vulnerabilities and protecting protecting the safety of the involved users.
\begin{figure}[htbp]
  \centering
  \includegraphics[width=0.5\textwidth]{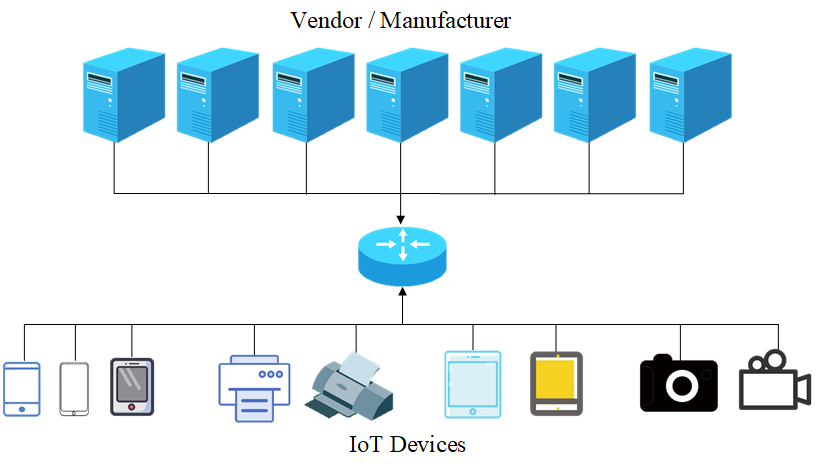}\\
  \caption{The Client-Server Architecture}\label{fig1}
\end{figure}
 Traditional software updates mainly based on the client-server architecture, as shown in Figure 1, create a single point of failure for denial of service (DoS) attacks. Delivering secure and reliable updates become a challenge issue for the vendors.

Building upon decentralization concept, the advent of blockchain technology may provide a solution for IoT \cite{khan}. Blockchain is a data structure that is based on hash functions that builds a linked list by using hash pointers. Each block stores the transactions in the peer-to-peer network. Some nodes are known as miners; they run the consensus algorithms such as proof of work (PoW) \cite{Nakamoto} to mine and generate a new block. The blockchain employs elliptic curve cryptography and SHA-256 hash function to provide security for data authentication and integrity. As a publicly verifiable ledger, it has a full history of the transaction and provides a global distributed trust.
Blockchain technology has widely applied to healthcare \cite{elb}, IoT \cite{su}, \cite{zhao}, and financial transactions \cite{and2014a} - \cite{chiesa2016} etc. There are several blockchain based solutions for IoT software and/or firmware updates. Several papers \cite{Du1,Du2,Du3,Du4,Du5,Du6,Du7,Du8,Du9} have studied related security issues.

\textbf{{Related Work.}}
Lee and Lee \cite{lee} proposed a secure firmware updates scheme for embedded devices in the IoT environments. They executed firmware checking and validation by using blockchain with a new block structure and the BitTorrent as a firmware sharing network for firmware download, to enhance availability and integrity of updates. Boudguiga et al. \cite{boudguiga} used the blockchain technology to ensure the availability and innocuousness of software updates. They added the trusted innocuousness nodes checking the integrity of updates and only the approved updates can be downloaded. Yohan et al. \cite{yohan} proposed a firmware update framework by utilizing PUSH-based firmware updates. They used smart contract and the consensus mechanism of blockchain to preserve the integrity of updates. Recently, Leiba et al. \cite{leiba} proposed decentralized incentivized delivery network for IoT software updates. The participating nodes of delivery network deliver update to IoT devices and the nodes can get the financial incentive from the vendors. However, these mechanisms are inadequate in the process of software updates for the privacy of the involved users. In certain circumstances, when a consumer buys a IoT device, his personal information could be automatically linked to the device. In the vehicle system, an on-board unit (OBU) embed into automatic vehicle as a sensing layer node. This node communicates with the roadside infrastructure and other peer vehicles. The leakage of user information can lead to privacy threats.

\textbf{{Contributions.}}
In this paper, we propose a new blockchain based privacy-preserving IoT software updates protocol. It not only protects the privacy of the updated IoT devices, but also delivers secure and reliable updates with an incentive mechanism. The proposed protocol utilizes blockchain, smart contract, double authentication preventing signature (DAPS) and outsourced attribute-based signature (OABS) to deliver secure and reliable updates. In this protocol, the vendor delivers the updates by using smart contract to provide a financial incentive to the transmission node that provides a proof-of-delivery that a single update was delivered to the IoT devices. A transmission node obtains proof-of-delivery by using double authentication preventing signature (DAPS) to carry out fair exchange. In the process of fair exchange, the transmission node exchanges an OABS of the IoT device by using DAPS. Then, it uses the OABS as s proof-of-delivery to receive the financial incentive. The main contributions of the proposed protocol are as follows:

 We propose a new concrete OABS scheme and prove the existential unforgeability under chose message attacks.

 We propose the system model and system component of a blockchain-based privacy-preserving IoT software updates protocol.

 We propose a concrete blockchian-based privacy-peserving IoT software updates protocol by integrating blockchian, smart contract, DAPS and our proposed OABS, which satisfies anonymity, proof-of-delivery unforgeability, fairness, authentication and integrity.

 We analyze the security requirements of a blockchian-based privacy-preserving IoT software updates protocol and provide security analysis of the proposed protocol.

 We implement the proposed blockchian-based privacy-preserving IoT software updates protocol using smart contract to demonstrate the validity of the proposed protocol.

\subsection{Organization:}
The remain of this paper is organized as follows. The model of blockchain based privacy-preserving software updates protocol is in Section \ref{sec2}.  The introduction of building blocks is given in Section \ref{sec3}. The details of blockchain based privacy-preserving software updates protocol and the security analysis and evaluation are described in Section \ref{sec4} and Section \ref{sec5}. Finally, we conclude the paper in Section \ref{sec6}.

\section{Blockchain based privacy-preserving software updates model}\label{sec2}
In the section, we introduce the blockchain based privacy-preserving software updates model and relevant security requirements.
\subsection{Blockchain based privacy-preserving software update model}
\begin{figure}[h]
  \centering
  \includegraphics[width=0.8\textwidth]{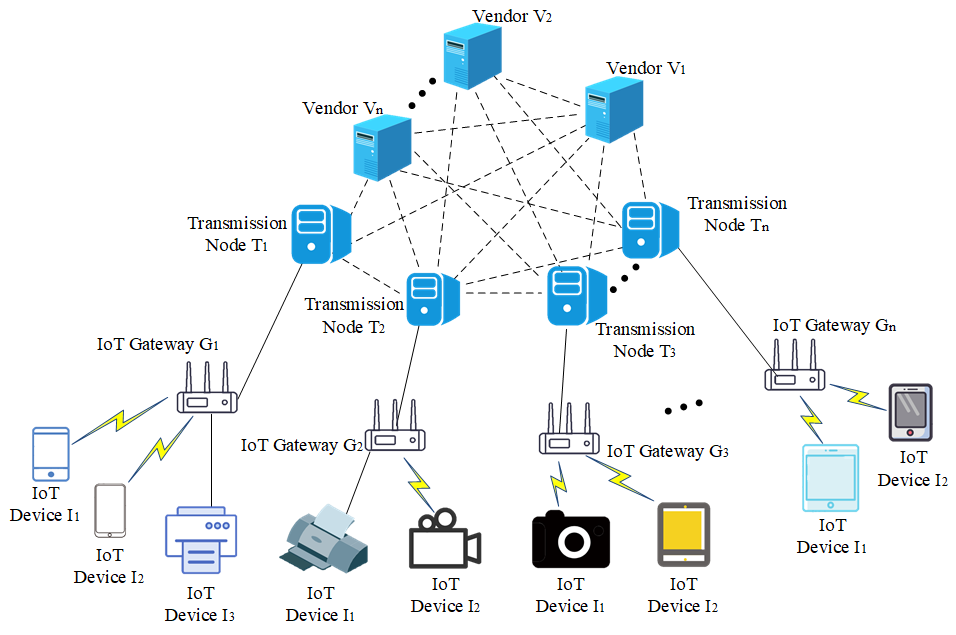}\\
  \caption{Blockchain-based privacy-preserving software updates model}\label{fig2}
\end{figure}

As shown in Figure \ref{fig2}. there are four participants including vendors, transmission nodes, IoT gateways and IoT devices.

\textbf{{Vendors:}} In blockchain network, the vendor as the provider of IoT devices to publish secure and reliable update.
It creates smart contract into blockchain network to provide financial incentive to the transmission node which delivers a single update to IoT devices. It acts as miner and verifies all transactions in the blockchain network. A set of vendors is denoted as $V=\{v_{1}, v_{2}, \cdot\cdot\cdot, v_{n}\}$, where $v_{i}\in V$.

\textbf{{Transmission nodes:}} The transmission node acting as broker or serve provider competitively finds targets and delivers the updates to IoT devices to obtain financial incentive. It acts as miner to maintain the blockchain network. The transmission nodes is denoted by $T=\{t_{1}, t_{2}, \cdot\cdot\cdot, t_{n}\}$, where $t_{i}\in T$.

\textbf{{IoT gateways:}} The IoT gateway acting as routing node such as WiFi router connects the IoT devices and it transmits the updates to IoT devices. It assists the IoT device to update and compute. A set of IoT gateways is denoted as $G_{gateway}=\{g_{1}, g_{2}, \cdot\cdot\cdot, g_{n}\}$, where $g_{i}\in G_{gateway}$.

\textbf{{IoT devices:}} The IoT devices are physical devices such as embedded device and smartphone. The IoT devices connected to a IoT gateway are denoted as a set $I=\{i_{1}, i_{2}, \cdot\cdot\cdot, i_{n}\}$, where $i_{i}\in I$.

\textbf{Definition 1.} (Blockchain based privacy-preserving software updates model). It includes a tuple ($Setup$, $KeyGen$, $Register$, $Publish$, $Query$, $Notification$, $Sign_{1}$, $Sign_{2}$, $Receive$) of polynomial time algorithms, which are defined as follows:
\begin{itemize}
\item \emph{$(params)\leftarrow Setup(\kappa)$}: This algorithm takes a security parameter $\kappa$ as input and it outputs the public parameters $params$.
\item \emph{$(pk,sk)\leftarrow KeyGen(params)$}: This algorithm takes $params$ as input and it outputs a public key $pk$ and a secret key $sk$.
\item \emph{$(L)\leftarrow Register(params, pk)$}: This algorithm takes $params$ and public key $pk$ as input and it outputs a register list $L$.
\item \emph{$(T_{p})\leftarrow Publish(params,update,sk)$}: This algorithm takes $params$, $update$ and $sk$ as input and it outputs the transaction $T_{p}$.
\item \emph{$(1/0)\leftarrow Query(params, update)$}: This algorithm takes $params$ and $update$ as input and it outputs a bit $b\in \{0, 1\}$.
\item \emph{$(1/0)\leftarrow Notification(T_{p}, update)$}:  This algorithm takes $T_{p}$ and $update$ as input and it outputs a bit $b'\in \{0, 1\}$. 
\item \emph{$(\sigma_{1})\leftarrow Sign_{1}(params, update, sk)$}: This algorithm takes $params, update$ and $sk$ as input and it outputs a signature $\sigma_{1}$.
\item \emph{$(\sigma_{2})\leftarrow Sign_{2}(params,\sigma_{1}, sk)$}: This algorithm takes $params, \sigma_{1}$ and $sk$ as input and it outputs a signature $\sigma_{2}$.
\item \emph{$(T_{r})\leftarrow Receive(params,\sigma_{2}, sk)$}: This algorithm takes $params, \sigma_{2}$ and $sk$ as input and it generates a signature $\sigma_{3}$. Then, it outputs a transaction $T_{r}$.
\end{itemize}
\subsection{Security requirements}
In the blockchain based privacy-preserving software updates model, it should satisfy the security requirements as follows.

\textbf{Completeness}: The completeness says that if the protocol is properly executed at all epochs, then an honest transmission node can get financial incentive and an honest vendor can distribute the updates to its IoT devices.


\textbf{Anonymity}: The protocol protects the privacy of the IoT devices. The IoT devices execute the protocol for updates without revealing the real identity of user.

\textbf{Proof-of-delivery Unforgeability.} The transmission node cannot claim to possession proof-of-delivery that he has not been provided.

\textbf{Fairness}:  The fairness is said that, either the transmission node obtains financial incentive and the vendor distributes the updates to its IoT device or neither the transmission node and the vendor get nothing, at the end of protocol.

\textbf{{Authentication and Integrity.}} For a new version update of the vendor, it should include a valid signature of the vendor to guarantee the authentication and integrity of updates.

\section{Building blocks}\label{sec3}
In this section, we review the smart contract and the cryptography algorithms used in the protocol.
\subsection{Smart contract}
Smart contract was firstly proposed by Nick Szabo in 1994 \cite{smart}. It is a computerized transaction protocol \cite{nick} and it has been applied to Bitcoin network, Ethereum \cite{wood2014} platform. In Bitcoin network, smart contract evolved from the scripting language of Bitcoin and it is a based on stack and not turning-complete language. As the next generation cryptocurrency and decentralized application , Ethereum \cite{wood2014} supports the smart contract, it offers more expressive expressions with turning-complete languages. Solidity \footnote{http://solidity.readthedocs.io/en/latest.}, a JavaScript-like languages, is the most widely adopted language for developing smart contract in Ethereum.  

\subsection{Double authentication preventing signatures}
In 2014, Poettering and Stebila \cite{poe1}, \cite{poe} proposed the concept of double authentication preventing signature (DAPS), which is a factoring-based setting and prevents compelled certificate creation attack. Later, Ruffing et al. \cite{ruffing} gave a construction of DAPS in the discrete logarithm setting and it is based on Merkle trees and chameleon hash functions. The DAPS was used to penalize the double spending of transactions. 
  We adopt the DAPS to blockchain based privacy-preserving software updates for fair exchange. We use the practical instantiation of DAPS scheme in the discrete logarithm setting \cite{derler}. Let $\Sigma_{DAPS}=(DAPS.Setup, DAPS.Kgen, DAPS.Sign, DAPS.Ver,$  $DAPS.Ext)$ be the DAPS scheme.
The DAPS is existential unforgeability under chosen message attacks $(\mathrm{EUF-CMA})$ secure in the random oracle model \cite{derler}. As a building block, it can be replaced by other double authentication preventing signature such as the post-quantum instantiation against quantum computer attacks and interested readers can refer to \cite{derler1}.

\subsection{Outsourced Attribute-based Signatures}
The concept of ABS was introduced by Maji et al. \cite{maji}, \cite{maji1}. In an ABS, a signer signs the message based on attributes satisfy the predicate and it doesn't revealing the identity of signer. It mainly applies to fine-grained access control such as anonymous authentication systems. 
In order to reduce bandwidth and computational overhead at the signer side, Chen et al. \cite{chen} proposed the efficient outsourced ABS (OABS), a signing-cloud service provider (S-CSP) assists signer to carry out computation. An OABS scheme includes five polynomial time algorithms: $OABS.Setup, OABS.KeyGen, OABS.Sign_{out}, OABS.Sign$, $OABS.Ver$.

\textbf{OABS.Setup:} This algorithm takes security parameter $\kappa$ and the attribute universe $U$ as inputs. It outputs the public parameter $params$ and the master key $MSK$.

\textbf{OABS.KeyGen:} This algorithm takes the public parameter $params$, the master key $MSK$ and an access structure $(\mathbb{M},\rho)$ as input and it outputs the outsourcing key $SK_{\mathbb{M},\rho}$ and the signing key $SK_{OABS}$.

\textbf{$\mathbf{OABS.Sign_{out}:}$} This algorithm takes the outsourcing key $SK_{\mathbb{M},\rho}$ and the authorized signing attribute set $W$ as input and it outputs the outsourced signature $\Gamma'$.

\textbf{OABS.Sign:} This algorithm takes a message $M$, the signing key $SK_{OABS}$ and the outsourced signature $\Gamma'$ as input and  outputs a signature $\Gamma$.

\textbf{OABS.Ver:} This algorithm takes the message $M$, the signature $\Gamma$ and the attribute set $W$ as inputs and it outputs a bit $b\in \{0, 1\}$.




\section{Blockchain based privacy-preserving IoT software updates protocol}\label{sec4}
\subsection{Overview}
 The privacy-preserving IoT software updates protocol works as follows. The vendor as one provider of the IoT devices initializes the system parameters. It maintains a list of its IoT devices and burns the secret key of device into the manufactured IoT devices. The transmission node registers with the vendor to deliver updates to the IoT devices and obtains financial incentive. Then, the vendor publishes updates by using smart contract and it commits to provide financial incentive to the transmission node that provides proof-of-delivery. The transmission node queries to download the updates that encrypted by the public key of the transmission node and it sends notification to the IoT gateways. Then, the IoT gateway checks the connected IoT devices to match the updates.
 The transmission node sends the ciphertext of updates with a DAPS to the IoT gateways. Then, the IoT gateway verifies the DAPS and sends the ciphertext of updates to the IoT devices. The IoT device generates OABS to the transmission node. When the transmission node receives an OABS, it generates a new DAPS. As a proof-of-delivery, it sends the DAPS and the OABS to blockchain network to receive the incentive. The IoT gateway extracts the secret key of the corresponding public key about the transmission node by using the extract algorithm of DAPS. It sends the secret key to the IoT device and the IoT device decrypts the ciphertext of updates. 
  We assume the IoT gateway is honest to the IoT device and does not collude with the transmission node. In blockchain network, each entity has the ECDSA key pair $(PK,SK)$, where $Sign_{SK}(m)$ denotes the ECDSA signature on message $m$.

\subsection{The details of protocol}
\begin{figure}[htbp]
  \centering
  \includegraphics[width=1.0\textwidth]{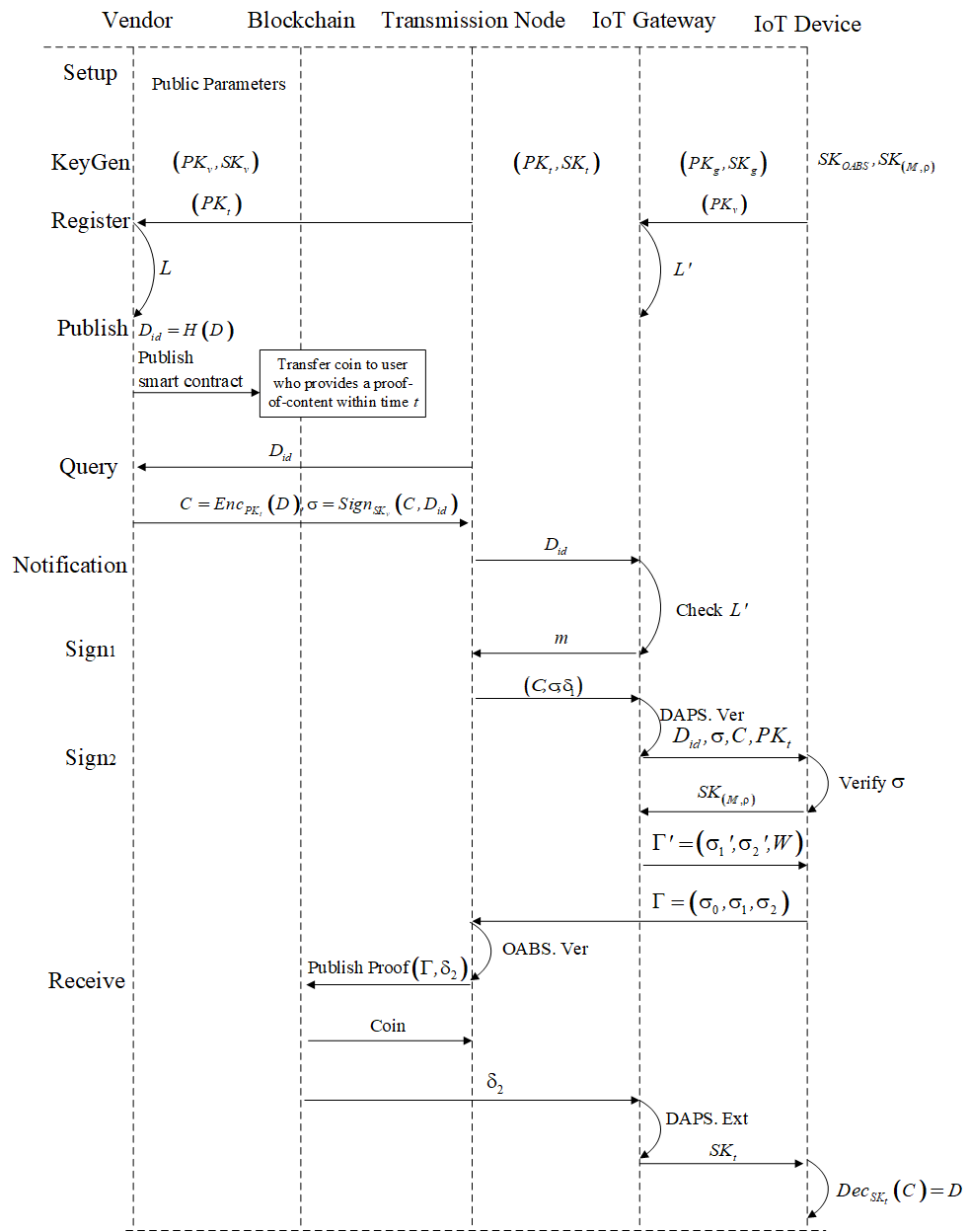}\\
  \caption{The sketch of the proposed protocol}\label{fig3}
\end{figure}

  See the Figure \ref{fig3}, the privacy-preserving IoT software update protocol sketch.
   Our OABS scheme is based on the ABS scheme of Rao \cite{rao}. 
     The concept of the computational $n$-Diffie-Hellman exponent problem, access structure and linear secret sharing scheme can refer to \cite{rao},\cite{waters}. We adopt ElGamal \cite{elgamal} public key encryption algorithm to encrypt data ($Enc$ denotes the ElGamal encryption algorithm and $Dec$ denotes the ElGamal decryption algorithm on data $m$). The concrete construction of the protocol is as follows.
\begin{itemize}
\item \emph{$Setup$}: The vendor runs OABS.Setup, it inputs a security parameter $\kappa$ and it outputs the bilinear paring $\Phi= ( q, G, G_{T}, e)$, where $G, G_{T}$ are cyclic multiplicative groups with order $q$. Let $\mathcal{M}=\{0,1\}^{l}$ be the message space. The attribute universe $U\in Z^{*}_{q}$ and one default attribute $\theta\in Z^{*}_{q}$. Let $n$ be a max size of the attribute set. It selects $\alpha\in Z_{q}$, a generator $g\in G$ and sets $Z=e(g,g)^{\alpha}$. $H:\{0,1\}^{*}\rightarrow \{0,1\}^{l}$ is a collision resistant hash function. Then, picks $V_{0},V_{1},\cdot\cdot\cdot,V_{n},u_{0},u_{1},\cdot\cdot\cdot,u_{l}\in G$. The master key is $MSK=\alpha$. Then, it calls DAPS.Setup to generate common reference string $crs$.

\item \emph{$KeyGen$}: The vendor generates an ECDSA key pair $(PK_{v},SK_{v})$. Then, it runs OABS.KeyGen to generate the secret key of the IoT devices for LSSS access structures $(\mathbb{M},\rho)$. Each row $i$ of the matrix $\mathbb{M}$ of size $l_{s}\times k_{s}$ is associated with an attribute $\rho(i)\in Z^{*}_{q}$. Then, it randomly chooses $\alpha_{1}\in Z^{*}_{q}$ such that $\alpha_{2}=\alpha-\alpha_{1}$ 
and computes the sharing $\{\lambda_{\rho(i)}=\mathbf{M}_{i} \mathbf{v}:i\in[l_{s}]\}$, where $\mathbf{M}_{i}$ is the $i$th row of $\mathbb{M}$ and the $\mathbf{v}\in Z^{k_{s}}_{q}$ such that $\mathbf{v}\mathbf{1}=\alpha_{1}, \mathbf{1}=(1,0,\cdot\cdot\cdot,0)$ is $k_{s}$ length vector. For each $i\in[l_{s}]$, the vendor chooses $r_{i}\in Z_{q}$ and computes $d_{i}=g^{\lambda_{\rho(i)}}V_{0}^{r_{i}}$,$d'_{i}=g^{r_{i}}$,$d''_{i}=\{d''_{i,x}:d''_{i,x}=(V_{1}^{-\rho(i)^{x-1}}V_{x})^{r_{i}}, \forall x=2,\cdot\cdot\cdot,n\}$. For default attribute $\theta$, the vendor chooses $r_{\theta}\in Z_{q}$ and computes $d_{\theta}=g^{\alpha_{2}}V_{0}^{r_{\theta}}$, $d'_{\theta}=g^{r_{\theta}}$, $d''_{\theta}=\{d''_{\theta,x}:d''_{\theta,x}=(V_{1}^{-\theta^{x-1}}V_{x})^{r_{\theta}}, \forall x=2,\cdot\cdot\cdot,n\}$.
    Finally, the vendor returns the outsourcing key $SK_{\mathbb{M},\rho}=\langle\{d_{i},d'_{i},d''_{i}:i\in [l_{s}]\}\rangle$, and the private key of IoT device $SK_{OABS}=(SK_{\mathbb{M},\rho},d_{\theta},d'_{\theta},d''_{\theta})$. Then, it burns $SK_{OABS}, SK_{\mathbb{M},\rho}$ and $PK_{v}$ into the manufactured IoT device. The transmission node calls DAPS.Kgen to generate its key pair $(pk_{t}, sk_{t})$, as well as generates an ECDSA key pair $(PK_{t},SK_{t})$ and the IoT gateway generates an ECDSA key pair $(PK_{g},SK_{g})$.

 \item \emph{\textbf{$Register$}:} The transmission node registers with the vendor and it sends $pk_{t}$ to the vendor. Then the vendor maintains a list $L$ which records the public key $pk_{t}$ of the transmission node. The IoT device sends $pk_{v}$ to the IoT gateway and the IoT gateway maintains a list $L'$ which records the public key $pk_{v}$.

\item \emph{$Publish$}: The vendor generates a update denoted as $D$ and sets $D_{id}=H(D)$. It publishes a smart contract to the blockchain network to provide financial incentive to the transmission node. As shown in Table \ref{ree}. the pseudocode of the smart contract. The vendor sets the limitation time $t$ as time epoch.
\begin{table*}[htbp]
  \begin{center}
     \begin{tabular}{|l|}\hline
     contract ProofOfDelivery\\
     \ \ \ function  ProofOfDelivery (v, t, $D_{id}$, n, W, L, x)\\
     \ \ \ \ \ \ owner $\leftarrow$ v\\
     \ \ \ \ \ \ limitationTime $\leftarrow$ t \\
      \ \ \ \ \ \ update $\leftarrow$ $D_{id}$ \\
     \ \ \ \ \ \ publicKeyList $\leftarrow$ L \\
     \ \ \ \ \ \ attributeSet $\leftarrow$ W \\
     \ \ \ \ \ \ counterUpdatedDevice $\leftarrow$ n-1 \\
          \ \ \ \ \ \ incentive $\leftarrow$ x \\
     \ \ \ \ \ \ balance $\leftarrow$ value \\
     \ \ \ function FinancialIncentive(OABS.Sign, DAPS.Sign, $pk_t$, $PK_{t}$)\\
     \ \ \ \ \ \ assert current time $t_{1}\leq$ limitationTime\\
    \ \ \ \ \ \ if OABS.Ver(attributeSet, OABS.Sign, update)\\
     \ \ \ \ \ \ if DAPS.Ver($pk_t$, DAPS.Sign) \\
     \ \ \ \ \ \ transfer(balance-incentive$\ast$counterUpdatedDevice, $PK_{t}$)\\
     \ \ \ \ \ \ counterUpdatedDevice = counterUpdatedDevice - 1\\
     \ \ \ function Withdraw()\\
      \ \ \ \ \ \ assert current time $t_{1}>$ limitationTime\\
       \ \ \ \ \ \ transfer (balance, owner)\\\hline
   \end{tabular}
  \end{center} \caption{The pseudocode of the smart contract} \label{ree}
\end{table*}

\item \emph{$Query$}: The transmission node queries the binary files of update $D_{id}$ and the vendor responses corresponding data. It encrypts the update $D$ with $pk_{t}$ to generate $C=Enc_{pk_{t}}(D)$ and $\sigma=Sign_{SK_{v}}(C,D_{id})$. Then, it sends $(C, \sigma)$ to the transmission node. The transmission node verifies the signature $\sigma$ and obtains the update $D$.

\item \emph{$Notification$}: The transmission node sends notification of a new update $D_{id}$ to the IoT gateway. Then the IoT gateway checks connected IoT devices and queries the updates by sending a random message $m$ to the transmission node.

\item \emph{$Sign_{1}$}: The transmission node calls DAPS.Sign to generate a signature $\delta_{1}$ about $(m, pk_{t})$. It sends $(C, \sigma, \delta_{1})$ to the IoT gateway. The IoT gateway calls DAPS.Ver to verify the signature $\delta_{1}$ and sends $(D_{id}, C, \sigma, pk_{t})$ to the IoT device.

\item \emph{$Sign_{2}$}: The IoT device verifies the signature $\sigma$. Then, it generates OABS to the transmission node. First, it sends outsouring key $SK_{\mathbb{M},\rho}$ to the IoT gateway and requests a partial signature. 
    The IoT gateway calls $OABS.Sign_{out}$ with the outsouring key $SK_{\mathbb{M},\rho}$ as follows.

  It obtains $\{w_{i}\in Z_{q}:i\in [l_{s}]\}$, where $I=\{i\in [l_{s}]:\rho(i)\in W\}$ such that $\sum_{i\in I}w_{i}\mathbf{M}_{i}=1$. Then, computes the coefficients $c_{1},c_{2},\cdots, c_{n}$ of the polynomial below.
  $$P(X)=\prod_{w\in W\cup\{\theta\}}(X-w)=\sum^{|W|+2}_{j=1}c_{j}\cdot X^{j-1}=\sum^{n}_{j=1}c_{j}\cdot X^{j-1}$$
  Set $c_{|W|+3}=\cdots=c_{n}=0$.
  It picks $r\in Z_{q}$ and computes $$\sigma'_{1}=g^{r}\prod_{i\in I}(d'_{i})^{w_{i}},\ \   \sigma'_{2}=(\prod_{i\in I}(d_{i}\prod^{n}_{x=2}(d''_{i,x})^{c_{x}})^{w_{i}})\cdot (V_{0}\prod_{k\in [n]}V_{k}^{c_{k}})^{r}$$
Then, it outputs the partial signature $\Gamma'=(\sigma'_{1},\sigma'_{2},W)$ to the IoT device.
After receiving the partial signature $\Gamma'=(\sigma'_{1},\sigma'_{2},W)$, the IoT device uses $SK_{OABS}$ to run the OABS.Sign algorithm. First, it computes $\sigma_{1}=\sigma'_{1}\cdot d'_{\theta}$ and $(m_{1},\cdots, m_{l})= H(pk_{t}||\sigma_{1}||W||\theta)$. It chooses $s\in Z_{q}$ and computes $\sigma_{0}=g^{s}$. Then, the IoT device computes the coefficients $c_{1},c_{2},\cdots, c_{n}$ of the polynomial as well as the outsourced signing algorithm. It computes $\sigma_{2}=d_{\theta}\prod^{n}_{x=2}(d''_{\theta,x})^{c_{x}}\cdot\sigma'_{2}\cdot (u_{0}\prod^{l}_{j=1}u_{j}^{m_{j}})^{s}$. Finally, outputs the signature 
$\Gamma=(\sigma_{0},\sigma_{1},\sigma_{2})$. Then, the IoT devices sends $\Gamma=(\sigma_{0},\sigma_{1},\sigma_{2})$ to the transmission node.

\item \emph{$ Receive$}: The transmission node runs OABS.Ver algorithm. It computes the coefficients $c_{1},c_{2},\cdots, c_{n}$ of the polynomial as well as the outsourcing signing algorithm and computes $(m_{1},\cdots, m_{l})= H(pk_{t}||\sigma_{1}||W||\theta)$. it verifies the equation $$Z\overset{?}{=}\frac{e(\sigma_{2},g)}{e(\sigma_{0},u_{0}\prod^{l}_{j=1}u_{j}^{m_{j}})e(\sigma_{1},V_{0}\prod_{k\in [n]}V_{k}^{c_{k}})}$$ Then, the transmission node calls DAPS.Sign to generate a new DAPS $\delta_{2}$. It calls smart contract to output a receive transaction $T_{r}$. Once the transaction $T_{r}$ is included in blockchain, the IoT gateway uses $\delta_{1}$ and $\delta_{2}$ to extract the secret key $sk_{t}$ corresponding to the public key $pk_{t}$ and sends $sk_{t}$ to the IoT device. Then the IoT device utilizes $sk_{t}$ to decrypt the ciphertext $C$ to obtain the updates $D$.

\end{itemize}


\section{Security and implementation}\label{sec5}
In this section, we analyze the security of the blockchain based privacy-preserving IoT software update protocol, then report the performance of the protocol.

\subsection{Security analysis}
The security of proposed protocol is guaranteed by following lemmas.

\textbf{Lemma 1.} The proposed blockchain based privacy-preserving IoT software updates protocol satisfies completeness.

\textbf{Proof.} The protocol is properly executed at all epochs. The vendor initializes system parameters. Then, it publishes smart contract to blockchain network for a new update. An honest transmission node can obtain financial incentive by publishing a proof-of-delivery to blockchain. By the proof-of-delivery the honest vendor can be sure that an update has been distributed to its IoT devices and an honest IoT device obtains the updates.
In the OABS scheme, for the attribute $\rho(i)\in W$, hence $0=P_{w\in W}(\rho(i))=\sum^{n}_{k=1}c_{k}\cdot\rho(i)^{k-1}$. We have $c_{1}=-\sum^{n}_{k=2}c_{k}\cdot\rho(i)^{k-1}$. The default attribute $\theta$ is same. Since $\sum_{i\in I}w_{i}\mathbf{M}_{i}=\mathbf{1}$, we have $\sum_{i\in I}\lambda_{\rho(i)}w_{i}=\alpha_{1}$. Now
\begin{eqnarray*}
 \sigma_{2} &= & d_{\theta}\prod^{n}_{x=2}(d''_{\theta,x})^{c_{x}}\cdot\sigma'_{2}\cdot(u_{0}\prod^{l}_{j=1}u_{j}^{m_{j}})^{s}\\
 &=& g^{\alpha_{2}}V_{0}^{r_{\theta}}\prod^{n}_{x=2}(V_{1}^{-\theta^{x-1}}V_{x})^{r_{\theta}c_{x}}\cdot(\prod_{i\in I}(d_{i}\prod^{n}_{x=2}(d''_{i,x})^{c_{x}})^{w_{i}})\cdot (V_{0}\prod_{x\in [n]}V_{x}^{c_{x}})^{r}\cdot (u_{0}\prod^{l}_{j=1}u_{j}^{m_{j}})^{s} \\
   &=& g^{\alpha_{2}}V_{0}^{r_{\theta}}\prod^{n}_{x=1}(V_{x}^{c_{x}})^{r_{\theta}}\cdot(\prod_{i\in I}(d_{i}\prod^{n}_{x=2}(d''_{i,x})^{c_{x}})^{w_{i}})\cdot (V_{0}\prod_{x\in [n]}V_{x}^{c_{x}})^{r}\cdot (u_{0}\prod^{l}_{j=1}u_{j}^{m_{j}})^{s}\\
 &=& g^{\alpha_{2}}(V_{0}\prod^{n}_{x=1}V_{k}^{c_{k}})^{r+r_{\theta}}\cdot(\prod_{i\in I}(g^{\lambda_{\rho(i)}}V_{0}^{r_{i}}\prod^{n}_{x=2}(V_{1}^{-\rho(i)^{x-1}}V_{x})^{r_{i}c_{x}})^{w_{i}})\cdot (u_{0}\prod^{l}_{j=1}u_{j}^{m_{j}})^{s}\\
 &=& g^{\alpha_{2}}(V_{0}\prod^{n}_{x=1}V_{k}^{c_{k}})^{r+r_{\theta}}\cdot g^{\sum_{i\in I}\lambda_{\rho(i)}w_{i}}(V_{0}V_{1}^{-\sum_{x=2}^{n}c_{x}\rho(i)^{x-1}}\prod^{n}_{x=2}V_{x}^{c_{x}})^{\sum_{i\in I}r_{i}w_{i}} \cdot (u_{0}\prod^{l}_{j=1}u_{j}^{m_{j}})^{s}\\
  &=& g^{\alpha_{2}}(V_{0}\prod^{n}_{x=1}V_{k}^{c_{k}})^{r+r_{\theta}}\cdot g^{\alpha_{1}}(V_{0}V_{1}^{c_{1}}\prod^{n}_{x=2}V_{x}^{c_{x}})^{\sum_{i\in I}r_{i}w_{i}} \cdot (u_{0}\prod^{l}_{j=1}u_{j}^{m_{j}})^{s}\\
  &=& g^{\alpha}(V_{0}\prod^{n}_{x=1}V_{k}^{c_{k}})^{r+r_{\theta}}\cdot (V_{0}\prod^{n}_{x=1}V_{x}^{c_{x}})^{\sum_{i\in I}r_{i}w_{i}} \cdot (u_{0}\prod^{l}_{j=1}u_{j}^{m_{j}})^{s}\\
  &=& g^{\alpha}(V_{0}\prod^{n}_{x=1}V_{k}^{c_{k}})^{r+r_{\theta}+\sum_{i\in I}r_{i}w_{i}}(u_{0}\prod^{l}_{j=1}u_{j}^{m_{j}})^{s}
\end{eqnarray*}
\begin{eqnarray*}
\sigma_{1} &= & \sigma'_{1}\cdot d'_{\theta}\\
&= & g^{r}\prod_{i\in I}(d'_{i})^{w_{i}}g^{r_{\theta}}\\
&= & g^{r}\prod_{i\in I}(g^{r_{i}})^{w_{i}}g^{r_{\theta}}\\
&= & g^{r+r_{\theta}+\sum_{i\in I}r_{i}w_{i}}
   \end{eqnarray*}
Therefore, $$Z=\frac{e(\sigma_{2},g)}{e(\sigma_{0},u_{0}\prod^{l}_{j=1}u_{j}^{m_{j}})e(\sigma_{1},V_{0}\prod_{k\in [n]}V_{k}^{c_{k}})}$$

 If $\delta_{1}$ and $\delta_{2}$ are valid DAPS, the IoT gateway can extract the secret key $sk_{t}$ of the corresponding public key $pk_{t}$ by running the DAPS.Ext algorithm. Then, it sends the secret key $sk_{t}$ to IoT device. The IoT device is able to decrypt the ciphertexts $C$ and gets updates by $D=Dec_{sk_{t}}(C)$.



\textbf{Lemma 2.} The proposed blockchain based privacy-preserving software update protocol satisfies anonymity.

\textbf{Proof.} In blockchain based privacy-preserving software update protocol, the anonymity of IoT device is derived from the OABS scheme. Here, we prove the OABS scheme satisfies signer privacy. For an OABS signature based on the message $pk_{t}$ with an attribute set $W$, it outputs the OABS form $\Gamma=(\sigma_{0},\sigma_{1},\sigma_{2})$, where $$\sigma_{0}=g^{s},\ \  \sigma_{1} = \sigma'_{1}\cdot d'_{\theta} = g^{r+r_{\theta}+\sum_{i\in I}r_{i}w_{i}},$$
 $$\sigma_{2} = d_{\theta}\prod^{n}_{x=2}(d''_{\theta,x})^{c_{x}}\cdot\sigma'_{2}\cdot(u_{0}\prod^{l}_{j=1}u_{j}^{m_{j}})^{s} = g^{\alpha}(V_{0}\prod^{n}_{x=1}V_{k}^{c_{k}})^{r+r_{\theta}+\sum_{i\in I}r_{i}w_{i}}(u_{0}\prod^{l}_{j=1}u_{j}^{m_{j}})^{s},$$
Let $\gamma=r+r_{\theta}+\sum_{i\in I}r_{i}w_{i}$, where $\sigma_{1}= g^{\gamma}, \sigma_{2}= g^{\alpha}(V_{0}\prod^{n}_{x=1}V_{k}^{c_{k}})^{\gamma}(u_{0}\prod^{l}_{j=1}u_{j}^{m_{j}})^{s}.$
The $(m_{1},\cdots, m_{l})= H(pk_{t}||\sigma_{1}||W||\theta)\in \{0,1\}^{l}$. $g$ is a random generator of $G$ and $V_{0},V_{1},\cdot\cdot\cdot,V_{n},u_{0},u_{1},\cdot\cdot\cdot,u_{l}\in G$ are public parameters. The $\alpha$ is master key, $s,\gamma$ are random exponents and $c_{1},c_{2},\cdots, c_{n}$ is the coefficients of the polynomial. Thus, the distribution of OABS $(\sigma_{0},\sigma_{1},\sigma_{2})$ is independent of the signing key, so the OABS scheme satisfies signer privacy.

\textbf{Lemma 3.} The proposed blockchain based privacy-preserving software update protocol satisfies proof-of-delivery unforgeability.

\textbf{Proof.} In the protocol, the proof-of-delivery unforgeability is derived from the OABS scheme. We prove the OABS scheme is existential unforgeability under selective-attribute attack and chosen message attacks secure. The proof follows from the ABS scheme of Rao et al.\cite{rao}, we give the description in Appendix A.

\textbf{Lemma 4.} The proposed blockchain based privacy-preserving software update protocol satisfies fairness.

\textbf{Proof.} First, we consider the vendor and the IoT device are malicious. As for a vendor, it distributes update to its IoT device without payment and the IoT device obtains the data of update without providing the OABS. Follow the protocol, the vendor encrypts the update $D$ with $pk_{t}$ to generate $C=Enc_{pk_{t}}(D)$ and $\sigma=Sign_{SK_{v}}(C,D_{id})$. Then, it sends $(C, \sigma)$ to the transmission node. The transmission node delivers the ciphertext $C$ to the IoT device. Without the OABS of the IoT device, the transmission node never submits its DAPS to blockchain. So, The IoT device is unable to get the secret key $sk_{t}$ to decrypt the ciphertext. The IoT device must send the OABS to the transmission node, it will get the secret key $sk_{t}$. The transmission node sends the DAPS and the OABS as proof-of-delivery to blockchain and obtains financial incentive from the vendor. We say that is contradiction that the vendor distributes update to its IoT device without payment and the IoT device obtains the data of update without providing the OABS. Thus, the probability of success for malicious the vendor and the IoT device is negligible.

 The other case is that the transmission node is malicious, the vendor and the IoT device are honest. The transmission node can get payment without submitting the DAPS. According to the smart contract, the miner can not verify the transaction $T_{r}$ without the DAPS, so the transmission node is unable to get financial incentive. In the limitation time $t$, the vendor can withdraw the payment. In this case, a malicious transmission node gets contradiction. The probability of success for malicious the transmission node is negligible. Therefore, The proposed blockchain based privacy-preserving software update protocol achieves fairness.



\textbf{Lemma 5.} The proposed blockchain based privacy-preserving software update protocol satisfies authentication and integrity.

\textbf{Proof.} For a new update of the vendor, it includes a valid ECDSA signature of the vendor $\sigma=Sign_{SK_{v}}(C,D_{id})$ with $D_{id}=H(D)$. The secure ECDSA signature guarantees the authentication and integrity of the update.

\subsection{Performance evaluation.} In the section, we implement our protocol to evaluate its performance. We refer to Solidity smart contract implemented on Ethereum. Since Ethereum does not provide the application programme interface (API) for OABS and DPAS, we will quantify the computation cost of cryptographic algorithms and the gas cost of smart contract separately.
We execute cryptographic algorithms by Miracl library \footnote{https://certivox.org/display/EXT/MIRACL}, and selects a CP elliptic curve for security level AES-80. The experiments platform are based on Dell (Windows 7 operation system with Intel(R) Core(TM) i5-2450M CPU 2.50 GHz and 4.00GB RAM). The average time cost of cryptographic algorithms with 1000 times is shown in Table \ref{2}.
 \begin{table*}[htbp]
  \begin{center}
  \begin{tabular}{cccccc}\hline
      Scheme& Setup & KGen& Sign/Enc& Ver/Dec& Ext \\\hline
       DAPS &0.007s& 0.013s&0.015s &0.031s &0.061s \\\hline
     ECDSA&0.006s& 0.002s& 0.004s& 0.010s&-\\\hline
    ElGamal &0.006s& 0.002s&0.011s &0.003s &- \\\hline

   \end{tabular}
  \end{center} \caption{The time cost of cryptographic algorithms on the laptop} \label{2}
\end{table*}
 Since the dominated computation of the IoT device is the signature of OABS, we evaluate the time cost of OABS signature algorithm. 
 When a IoT device owns 50 attributes, the time cost is almost 155ms. In the protocol, the total time cost for a IoT device is 168ms including OABS.Sign, ECDSA.Sign and Elgamal.Dec algorithm. This is an acceptable result for resource limited device.

We implement smart contract in Solidity with the Web3j and deploy smart contract to run different functions of the blockchain based privacy-preserving software updates protocol. The implementation of smart contract needs a few ether and the estimates of gas cost is provided in Table \ref{re}.
\begin{table*}
 \begin{center}
  \begin{tabular}{lcccc}\hline
    Function & Transaction Gas&Execute Gas & Gas cost(ether)  \\\hline
    ProofOfDelivery &445140&319096&0.01528472\\\hline
     FinancialIncentive& 22657 &1009& 0.00047332 \\\hline
     Withdraw &23027&1755&0.00049564\\\hline
   \end{tabular}
  \end{center} \caption{Gas cost of the smart contract} \label{re}
\end{table*}

\section {Conclusion}\label{sec6}
We describe a new blockchain based privacy-preserving IoT software update with proof-of-delivery protocol which utilizes blockchain, smart contract, double authentication preventing signature (DAPS) and outsourced attribute-based signature (OABS) to deliver secure and reliable update. It protects the privacy of IoT devices, as well as delivers secure and reliable update with an incentive mechanism. In this protocol, the vendor can deliver update to its IoT device by using smart contract. The transmission node can obtain financial incentive by providing a proof-of-delivery. We implemented smart contract in Solidity to demonstrate the validity of the proposed blockchain based privacy-preserving software update protocol. 

\textbf{Acknowledgement:} This work was supported by National Key R\&D Program of China (2017YFB0802000), National Natural Science Foundation of China (61872229, 61802239), National Cryptography Development Fund during the 13th Five-year Plan Period (MMJJ20170216), Fundamental Research Funds for the Central Universities(GK201702004, GK201803061, 2018CBLY006) and China Postdoctoral Science Foundation (2018M631121).



\begin{appendix}

\section{Unforgeability}
\textbf{Theorem 1.} (Unforgeability) Assume $n-CDHE$ problem is $(T,\epsilon)$ hard in $G$. Then, Our OABS scheme is $(T', q_{ok}, q_{k}, q_{s},\epsilon')$-EUF-sAtt-CMA secure.

\textbf{Proof.} Suppose that an adversary $\mathcal{A}$ can break $(T', q_{ok}, q_{k}, q_{s},\epsilon')$-EUF-sAtt-CMA security of our scheme, there will exist a simulator $\mathcal{S}$ that can solve the $n-CDHE$ problem with a non-negligible probability $\epsilon$ by using $\mathcal{A}$'s forgery.

The simulator $\mathcal{S}$ is given the hard problem instantiation, the parameters $(q,G,G_T,e)$ and $\overrightarrow{y}=(g, g^{\alpha},g^{\alpha^{2}}$ ,$\cdots,g^{\alpha^{n}}$ $,g^{\alpha^{n+2}}\cdots, g^{\alpha^{2n}})$, where $\alpha\in Z_{q}$ and $g\in G$ is a generator of $G$. Let $g_{i}=g^{\alpha^{i}}$ and denote $\Lambda=(\alpha,\alpha^{2},\cdots,\alpha^{n})$.

 \textbf{Init.} The simulator $\mathcal{S}$ specifies one default attribute $\theta$ and the attribute universe $U\in Z^{*}_{q}$, where $|U|=n, n$ is a bound on the size of attribute set. Then, the adversary $\mathcal{A}$ sends the challenge attribute set $W^{*}$ $(|W^{*}<n|)$ to $\mathcal{S}$.

 \textbf{Setup.} The simulator $\mathcal{S}$ selects a collision resistant hash function $H:\{0,1\}^{*}\rightarrow \{0,1\}^{l}$ and picks $\alpha',\alpha_{2}\in Z_{p}$. $\mathcal{S}$ sets \boxed{\alpha=\alpha_{1}+\alpha_{2}=(\alpha^{n+1}+\alpha')+\alpha_{2}} implicitly and sets $Z=e(g_{1},g_{n})\cdot e(g,g)^{\alpha'}\cdot e(g,g)^{\alpha_{2}}$. It picks $\gamma_{i}\in Z_{q}$ and sets $V_{i}=g^{\gamma_{i}}g_{i}$ for each $i\in [l_{s}]$. Then, it computes the coefficients $\mathbf{c^{*}}=(c^{*}_{1},c^{*}_{2},\cdots, c^{*}_{n})$ of the polynomial below.$$P_{W^{*}}(X)=\prod_{w\in W^{*}\cup\{\theta\}}(X-w)=\sum^{n}_{j=1}c^{*}_{j}\cdot X^{j-1}$$
 $\mathcal{S}$ selects $\gamma_{0}\in Z_{q}$ and sets $V_{0}=g^{\gamma_{0}}g_{1}^{-c_{1}^{*}}\cdots g_{n}^{-c^{*}_{n}}=g^{\gamma_{0}-\Lambda c^{*}}$. Then, $\mathcal{S}$ prepares $u_{0},u_{1},\cdot\cdot\cdot,u_{l}$. It picks a integer $\beta$ such that $\{0\leq \beta \leq l\}$. It selects $\mu_{0},\mu_{1},\cdots,\mu_{l}\in Z_{q}$, and $\delta_{0},\delta_{1},\cdots,\delta_{l}\in\{0,\cdots,\tau-1\}$, where $\tau=2q$($q$ is the number of signing queries). Then, it defines $u_{i}=g_{n}^{\delta_{i}}\cdot g^{\mu_{i}}$ for each $i\in [l]$ and $u_{0}=g_{n}^{-\beta\cdot\tau+\delta_{0}}g^{\mu_{0}}$. Then, it defines two function $F(\mathbf{m})$ and $J(\mathbf{m})$ for $\mathbf{m}=(m_{1},\cdots, m_{l})$, where $F(\mathbf{m})=\delta_{0}+\Sigma_{i\in [l]}m_{i}\delta_{i}-\beta\cdot\tau$ and $J(\mathbf{m})=\mu_{0}+\Sigma_{i\in [l]}m_{i}\mu_{i}.$ Finally, $\mathcal{S}$ sends the public parameters $params = (\Phi, U, n, g, Z, V_{0}, V_{1}, \cdot\cdot\cdot, V_{n}, u_{0}, u_{1}, \cdot\cdot\cdot , u_{l}) $ to $\mathcal{A}$.

 \textbf{KeyGen Query.} The adversary $\mathcal{A}$ makes outsourcing key query and signing key query as follows.

 -\textbf{Outsourcing key query.} Upon receiving an outsourcing key request, the simulator $\mathcal{S}$ performs simulation as follows. $\mathcal{S}$ constructs the outsourcing key $SK_{\mathbb{M},\rho}$ for the LSSS access struture $(\mathbb{M},\rho)$ with the $W^{*}$ does not satisfy $(\mathbb{M},\rho)$. Each row $i$ of the matrix $\mathbb{M}$ of size $l_{s}\times k_{s}$ denoted $\mathbf{M}_{i}$ is associated with an attribute $\rho(i)\in Z^{*}_{q}$. Since $W^{*}$ does not satisfy $(\mathbb{M},\rho)$, there is a vector $\mathbf{v_{1}}\in Z_{q}$ for $\mathbf{v_{1}}\mathbf{1}=-1$ and $\mathbf{M_{i}v_{1}}=0$ for $\{\rho(i)\in W^{*}:i\in [l_{s}]\}$, and $\mathbf{1}=(1,0,\cdots,0)$ is $k_{s}$ length vector. $\mathcal{S}$ selects a vector $\mathbf{v_{2}}$ such that $\mathbf{v_{2}}\mathbf{1}=0$. Then, it sets \boxed{\mathbf{v}=-\alpha_{1}\mathbf{v_{1}}+\mathbf{v_{2}}=-(\alpha^{n+1}+\alpha')\mathbf{v_{1}}+\mathbf{v_{2}}} implicitly. Here $\mathbf{v}\mathbf{1}=-\alpha_{1}\mathbf{v_{1}}\mathbf{1}+\mathbf{v_{2}}\mathbf{1}=-(\alpha^{n+1}+\alpha')\mathbf{v_{1}}\mathbf{1}+\mathbf{v_{2}}\mathbf{1}=\alpha^{n+1}+\alpha'=\alpha_{1}$. $\mathcal{S}$ to simulate the outsourcing key as follows. For each $i\in [l_{s}]$, it has two cases.

 1). If $\rho(i)\in W^{*}$, then $\mathbf{M_{i}v_{1}}=0$. $\lambda_{\rho(i)}=\mathbf{M}_{i} \mathbf{v}=-(\alpha^{n+1}+\alpha')\mathbf{M_{i}}\mathbf{v_{1}}+\mathbf{M_{i}}\mathbf{v_{2}}=\mathbf{M_{i}}\mathbf{v_{2}}$. $\mathcal{S}$ picks $r_{i}\in Z_{q}$ and computes
 $d_{i}=g^{\lambda_{\rho(i)}}V_{0}^{r_{i}}$,$d'_{i}=g^{r_{i}}$,$d''_{i}=\{d''_{i,x}:d''_{i,x}=(V_{1}^{-\rho(i)^{x-1}}V_{x})^{r_{i}}, \forall x=2,\cdot\cdot\cdot,n\}$.

 2). If $\rho(i)\notin W^{*}$, then $P(X)=P(\rho(i))\neq 0$. $\sum^{n}_{j=1}c^{*}_{j}\cdot \rho(i)^{j-1}\neq 0$, So $\mathbf{c}^{*}\boldsymbol{\rho_{i}}\neq0$, where $\boldsymbol{\rho_{i}}=(1, \rho(i),\cdot\cdot\cdot, \rho(i)^{n-1})$. We have $\lambda_{\rho(i)}=\mathbf{M}_{i} \mathbf{v}=\mathbf{M_{i}}(\mathbf{v_{2}}-\alpha'\mathbf{v_{1}})-a^{n+1}\mathbf{M_{i}}\mathbf{v_{1}}$.
 $\mathcal{S}$ picks $r'_{i}\in Z_{q}$ and it sets $r_{i}=r'_{i}-\frac{\mathbf{M_{i}}\mathbf{v_{1}}}{\mathbf{c}^{*}\boldsymbol{\rho_{i}}}\Delta \boldsymbol{\rho_{i}}$ implicitly, where $\Delta=(\alpha^{n},\cdot\cdot\cdot, a)$. Then, it computes $d_{i}
 = g^{\lambda_{\rho(i)}}V_{0}^{r_{i}}$,
 $d'_{i}=g^{r_{i}}$,
 $d''_{i}=\{d''_{i,x}:d''_{i,x}=(V_{1}^{-\rho(i)^{x-1}}V_{x})^{r_{i}}, \forall x=2,\cdot\cdot\cdot,n\}$.
The simulator $\mathcal{S}$ returns the outsourcing key $SK_{\mathbb{M},\rho}=\langle \{d_{i},d'_{i},d''_{i}:i\in [l_{s}]\}\rangle$ to the adversary $\mathcal{A}$.

-\textbf{Sign key query.} Upon receiving a signing key request, $\mathcal{S}$ performs simulation as follows. $\mathcal{S}$ chooses an attribute $\theta$ and $r_{\theta}\in Z_{q}$. Then it uses \boxed{\alpha_{2}} to compute $d_{\theta}=g^{\alpha_{2}}V_{0}^{r_{\theta}}$, $d'_{\theta}=g^{r_{\theta}}$, $d''_{\theta}=\{d''_{\theta,x}:d''_{\theta,x}=(V_{1}^{-\theta^{x-1}}V_{x})^{r_{\theta}}, \forall x=2,\cdot\cdot\cdot,n\}$. Then, the simulator $\mathcal{S}$ returns the signing key $SK_{OABS}=(d_{\theta},d'_{\theta},d''_{\theta})$ to the adversary $\mathcal{A}$.

\textbf{Sign query.}  $\mathcal{A}$ makes signing query on a message $M$ with an attribute set $W$. $\mathcal{S}$ constructs a LSSS access struture $(\mathbb{M},\rho)$ such that the $W$ satisfies $(\mathbb{M},\rho)$. Then, it checks whether $W^{*}$ satisfy the LSSS access struture $(\mathbb{M},\rho)$. If not, $\mathcal{S}$ generates the outsourcing key $SK_{\mathbb{M},\rho}$ and the signing key $SK_{OABS}$ for keygen query phase. Then, It returns the signature $\Gamma=(\sigma_{0},\sigma_{1},\sigma_{2})$ to $\mathcal{A}$. If $W^{*}$ satisfies the LSSS access struture $(\mathbb{M},\rho)$. $\mathcal{S}$ randomly chooses $\delta'\in Z_{q}$, and sets $\sigma_{1}=g^{\delta'}$. Then, it computes $\mathbf{m}=(m_{1},\cdots, m_{l})= H(M||\sigma_{1}||W||\theta)\in \{0,1\}^{l}$. If $F(\mathbf{m})=0\ mod\ q$, then the simulation aborts. Otherwise, $\mathcal{S}$ computes the coefficients $\mathbf{c}=(c_{1},c_{2},\cdots, c_{n})$ by the polynomial $P_{W}(X)=\prod_{w\in W\cup\{\theta\}}(X-w)=\sum^{n}_{j=1}c_{j}\cdot X^{j-1}.$ $\mathcal{S}$ randomly chooses $s'$ to generate $\sigma_{0}=g^{s'}g_{1}^{-1/F(\mathbf{m})}$,
$\sigma_{2}=g^{\alpha'+\alpha_{2}}(V_{0}\prod^{n}_{x=1}V_{k}^{c_{k}})^{\delta'}$ $(u_{0}\prod^{l}_{j=1}u_{j}^{m_{j}})^{s'}g_{1}^{-J(\mathbf{m})/F(\mathbf{m})}.$ Then, $\mathcal{S}$ returns the signature $\Gamma=(\sigma_{0},\sigma_{1},\sigma_{2})$ to $\mathcal{A}$.

\textbf{Forgery.} The adversary $\mathcal{A}$ outputs a forgery $\Gamma^{*}=(\sigma_{0}^{*},\sigma_{1}^{*},\sigma_{2}^{*})$ on message $M^{*}$ with the attribute set $W^{*}$. $\mathcal{S}$ checks whether $\Gamma^{*}$ is valid and $(M^{*},W^{*})$ have never been queried. If it not, $\mathcal{S}$ aborts. If it holds, which means that $\sigma_{0}^{*}=g^{s}, \sigma_{1}^{*}=g^{\delta'},\sigma_{2}^{*}=g^{\alpha^{n+1}+\alpha'+\alpha_{2}}(V_{0}\prod^{n}_{x=1}V_{k}^{c_{k}})^{\delta'}(u_{0}\prod^{l}_{j=1}u_{j}^{m^{*}_{j}})^{s}$. Let $\boldsymbol{\gamma}=(\gamma_{1}, \cdot\cdot\cdot, \gamma_{n})$, and  $u_{0}\prod^{l}_{j=1}u_{j}^{m^{*}_{j}}=g_{n}^{F(\mathbf{m})}g^{J(\mathbf{m})}$ and $V_{0}\prod^{n}_{x=1}V_{k}^{c_{k}}=g^{\gamma_{0}+\boldsymbol{\gamma} \mathbf{c}^{*}}$. Then, it computes $\mathbf{m^{*}}=(m^{*}_{1},\cdots, m^{*}_{l})= H(M^{*}||\sigma^{*}_{1}||W^{*}||\theta)\in \{0,1\}^{l}$. If $F(\mathbf{m^{*}})\neq0\ mod\ q$, then the simulation aborts. If $F(\mathbf{m}^{*})=0\ mod\ q$,
 the simulator $\mathcal{S}$ computes

$\frac{\sigma^{*}_{2}}{g^{\alpha'}g^{\alpha_{2}}(\sigma_{0}^{*})^{J(\mathbf{m}^{*})}(\sigma^{*}_{1})^{\gamma_{0}+\boldsymbol{\gamma} \mathbf{c}^{*}}}
=\frac{g^{\alpha^{n+1}+\alpha'+\alpha_{2}}(V_{0}\prod^{n}_{x=1}V_{k}^{c_{k}})^{\delta'}(u_{0}\prod^{l}_{j=1}u_{j}^{m^{*}_{j}})^{s}}{g^{\alpha'}g^{\alpha_{2}}(g^{s})^{J(\mathbf{m}^{*})}(g^{\delta'})^{\gamma_{0}+\boldsymbol{\gamma} \mathbf{c}^{*}}}=g^{\alpha^{n+1}}$

\end{appendix}

\end{document}